\documentclass[12pt]{article}
\usepackage{a4wide}
\usepackage{amssymb}
\usepackage{graphicx}
\begin{document}
{\renewcommand{\thefootnote}{\fnsymbol{footnote}} \begin{center}
    {\LARGE  Relativistic implications of entropy and purity}\\
    \vspace{1.5em} Joseph Balsells\footnote{e-mail address: {\tt
        jab934@cornell.edu}\\ New address: Cornell Laboratory for
      Accelerator-based Sciences and Education, Cornell University, Ithaca,
      New York, 14853, USA} and Martin Bojowald\footnote{e-mail address: {\tt
        bojowald@psu.edu}}
\\
\vspace{0.5em}
Institute for Gravitation and the Cosmos,\\
The Pennsylvania State
University,\\
104 Davey Lab, 251 Pollock Road, University Park, PA 16802, USA\\
\vspace{1.5em}
\end{center}
}

\setcounter{footnote}{0}

\begin{abstract}
 A quantum object is extended by virtue of uncertainty.  When subjected to
  gravity, different parts of its wave function experience distinct local
  relativistic effects, leading to tidal and interference phenomena absent in
  the classical limit. These effects can be incorporated into a geometric
  extension of classical spacetime. For states quantum correlated in at least
  two directions, a complete description of motion requires a non-Riemannian
  geometry whose form is controlled by the state's entropy and purity and
  affects a broad range of phenomena from lab measurements to Hawking
  radiation. A specific implication of this framework is the appearance of
  quantum parameters in the time-dilation law in addition to the usual
  dependence on velocity and gravitational potential. The quantum-corrected
  time-dilation law is universal: the corrections depend solely on the
  external degrees of freedom and are independent of internal details of the
  clock mechanism.
\end{abstract}

\section{Introduction}

Relativistic quantum information \cite{QuantumInfoRel} aims to combine crucial
ingredients of relativity and quantum physics by taking into account space-time
properties such as frame transformations as well as non-classical statistical
effects. A popular subject is given by entanglement properties of quantum fields
on curved space-time. Here, we show that even the more tractable case of quantum
mechanics gives rise to non-trivial phenomena, based on novel methods that allow
us to rewrite Schr\"odinger evolution as geodesic motion on an extended
space-time manifold. The extension includes new dimensions parameterized by
quantum fluctuations and correlations, and we show that it carries a
non-Riemannian geometry controlled by entropy and purity. We identify a specific
correlation parameter that is subject to the new geometry, and derive its
relationship with second-order moments of the state.

Physical implications of geodesic motion can be seen in tidal forces and in time
dilation. A quantum object is extended due to quantum uncertainty and therefore,
in a heuristic semiclassical picture, experiences the gravitational field
averaged over an entire region even if it describes a point particle. Since its
wave function or density matrix provides the weights for this average, its
motion, effectively described by tidal forces, deviates from point-particle
motion in ways determined by its state parameters. In addition, its energy and
effective mass depend on the state, and therefore the object is subject to
state-dependent time dilation. If suitable control over the quantum state of an
atomic reference frame could be realized, then these effects could manifest as
measurable frequency drifts in the next generation of high-precision atomic clocks.

Classical geodesic motion is derived by extremizing the arc length given by the
curve integral of proper time,
\begin{equation}
  \label{ProperTimeg} \int \mathrm{d }\tau = \frac{1}{c}\int\sqrt{-g_{ab}(x)
\mathrm{d }x^a\mathrm{d }x^b}\,,
\end{equation} on Riemannian space-time with metric tensor $g_{ab}$. This
variational principle admits a Hamiltonian formulation in which the generalized
velocities $\dot{x}^a = {\rm d}x^a/{\rm d}\tau$ are replaced by
canonical momenta $p_a=mg_{ab}\dot{x}^b$ for a particle of mass $m$. The
corresponding Hamiltonian reads
\begin{equation}
  \label{H} H(x^a,p_a) = \frac{1}{2m} \left( g^{ab}p_a p_b + (mc)^2 \right)\,.
\end{equation}
and Hamilton’s equations derived from \(H\) are equivalent to the geodesic equation.
The additive constant is required for consistency with the mass-shell constraint
$H=0$, which in flat space-time reproduces the well-known
energy-momentum relation $E^2/c^2=|\vec{p}|^2+m^2c^2$.

Unlike in non-relativistic mechanics, the Hamiltonian $H$ vanishes on physical
trajectories and does not represent the energy of the particle, which is instead
given by $E/c=p^0=g^{0a}p_a$.  The Hamiltonian vanishes because there is no
absolute time coordinate in relativity.

In the Hamiltonian formulation of relativistic particle dynamics, the
Hamiltonian acts both as (i) the generator of evolution (via Hamilton's
equations) and (ii) a constraint that determines how the system’s energy
content governs the progression of time along its worldline.  This dual role
can be extended to the quantum domain without additional assumptions about how
gravity might couple to a wave function or density matrix: the classical
Hamiltonian is replaced directly with its expectation value
$\langle\hat{H}\rangle$ in a suitable state.

Since the state itself
is subject to evolution, it is important to keep track of such changes along
a geodesic. The same evolution also experiences time dilation as the quantum
object moves along its geodesic. Because these two effects are interrelated, it
is impossible to split the calculation in two, a quantum one for the wave
function or density matrix with respect to (proper) time, and a classical one
for the geodesic and time dilation in a curved manifold. Even if we do not
change the space-time geometry, considering the quantum object as a test
particle, there is a subtle back-reaction effect because the changing energy
expectation value of an evolving state affects the dilation of time with respect
to which it is evolving.

\section{Quantum and relativity unified}

The problem therefore requires a unified
treatment of relativistic and quantum physics that incorporates space-time
geometry. We now show that such a treatment is possible even in the absence of
quantum gravity. The key tool is a Hamiltonian and quasiclassical formulation
of quantum mechanics, which allows classical relativistic couplings to extend
smoothly into the perturbative quantum regime.  Consistency between the
spacetime and phase-space geometries ensures that the induced gravitational
couplings of quantum degrees of freedom are uniquely determined up to
canonical transformation.

In this framework, the energy expectation value $\langle\hat{E}\rangle$
depends not only on the classical phase-space variables $(x^a,p_b)$ but also
on additional quantum variables that characterize the quantum state’s internal
geometry through its fluctuations, correlations, and purity.  These additional
quantities behave as dynamical coordinates with their own conjugate momenta,
and so define an extended phase space.  The resulting
Hamiltonian description treats the evolution of the quantum state and the
relativistic geometry through which it propagates on equal
footing: both arise as manifestations of geodesic motion on a unified,
non-Riemannian geometry.

For a system with two classical degrees of freedom, $x_1,x_2$ and conjugate
momenta $p_1,p_2$, the effective Hamilton function to first order in $\hbar$
follows from an expansion of the energy expectation value around the mean values
of the basic operators
\begin{equation} \label{Hexp} \langle\hat{E}\rangle\approx
E(\langle\hat{x}_1\rangle,\langle\hat{p}_1\rangle,\langle\hat{x}_2\rangle,
\langle\hat{p}_2\rangle)+\frac{1}{2}
\sum_{i=1}^4\sum_{j=1}^4\frac{\partial^2E}{\partial y_i\partial y_j}
\Delta(y_iy_j)
\end{equation} where the second-order central moments
\begin{equation}
  \label{D2moments}
  \Delta(y_iy_j) :=\langle(\hat{y}_i-\langle\hat{y}_i\rangle)
  (\hat{y}_j-\langle\hat{y}_j\rangle)\rangle_{\rm symm}
\end{equation}
are taken in completely symmetric ordering, and $y_i\in\{x_1,p_1,x_2,p_2\}$.
For semiclassical states, such as Gaussians, higher-order moments contribute
only at higher orders in \(\hbar\) and can be neglected in a first treatment.

In the present context, evolution with respect to the coordinate time
$t$ is governed by the effective Hamiltonian obtained by applying this
expansion to the energy expression $E=cp^0$, where \(p^0\) is determined by solving the constraint
$H=0$ in Eq.~(\ref{H}). The explicit form of the resulting Hamiltonian is reproduced in
\cite{QuantumFinsler}. This Hamilton function generates coupled evolution of both the classical expectation values \((x^i, p_i)\), \(i=1,2\), and the quantum statistical variables \(\Delta(y_iy_j)\), consistently incorporating their mutual back-reaction.

The references \cite{Bosonize,EffPotRealize} showed that the quantum
coordinates \(\Delta(y_iy_j)\) close under a Poisson bracket derived from the
quantum commutator, forming a Lie
algebra isomorphic to \({\rm sp}(4,\mathbb{R})\). The same papers constructed a
mapping that replaces the statistical moments with canonical pairs
\begin{equation}
  \label{eq:5}
  (s_1, p_{s_1}), \;
  (s_2, p_{s_2}), \;
  (\alpha, p_\alpha), \;
  (\beta, p_\beta)
\end{equation}
together with two Casimir functions \(C_1\) and \(C_2\) that have vanishing
Poisson brackets with any other expression. These variables satisfy the canonical Poisson algebra
\begin{equation}
  \label{eq:6}
  \{ s_i, p_{s_j} \} = \delta_{ij},
  \enspace
  \{ \alpha, p_\alpha \} = 1,
  \enspace
  \{ \beta, p_\beta \} = 1
\end{equation}
with all other Poisson brackets vanishing. The explicit relationship between
these variables and the underlying statistical moments
is summarized in
the Appendix.

This canonical realization leads to the effective energy function
\begin{equation}
  \label{eq:E2dof-def}
  E_{\mathcal{Q}} =
  E_{\mathcal{Q}}(x_1,p_1,x_2,p_2,s_1,p_{s_1},s_2,p_{s_2},\alpha,p_{\alpha},\beta,p_{\beta};C_1,C_2)
\end{equation}
defined on an extended phase space with coordinates
$(x_1,x_2,s_1,s_2,\alpha,\beta)$ and conjugate momenta
$(p_1,p_2,p_{s_1},p_{s_2},p_{\alpha},p_{\beta})$, together with two conserved
quantities $C_1$ and $C_2$ fixed by the initial state.

This realization is unique up to canonical transformations and is determined
by the requirement that all coordinates remain functionally independent.
Consequently, the time derivative of each coordinate follows from the
Hamiltonian partial derivative with respect to its conjugate momentum,
ensuring the evolution preserves the canonical symplectic structure.

When this procedure is applied to the geodesic Hamiltonian~(\ref{H}),
Hamilton's equations jointly determine the geodesic motion (through the
classical coordinates) and the evolution of the quantum state along that
geodesic.  The two are linked by relativistic interference effects, so that
changes in the quantum state feed back into the effective notion of time
experienced along its trajectory.

This interplay is summarized by the quantum proper time functional,
\begin{equation} \label{ProperTime} \Delta\tau=\int
\sqrt{1-\frac{2}{mc^2}\langle\hat{H}\rangle}\;{\rm d}\tau
\end{equation}
integrated along the geodesic.  In this expression, \(\langle\hat{H}\rangle\)
is related to \(\langle\hat{E}\rangle\) by the inverse of the procedure that
determined \(E = c p^0\) from \(H\). As a  result, the accumulated proper
time depends not only on the classical motion but also on the evolving quantum
state through its expectation value of the Hamiltonian.

\section{Canonical quantum information}

The variables \(x_1\) and \(x_2\)
represent independent position coordinates of the quantum system.  In the
covariance-matrix representation, these degrees of freedom act formally
as canonical modes with conjugate quadratures
\((x_i,p_i)\). Although the covariance-matrix framework originated for Gaussian optical states, it more generally characterizes the quadratic Hamiltonian (second-moment) dynamics of both Gaussian and non-Gaussian states. This correspondence allows the second-order relativistic
quantum mechanics developed here to be recast in the language of continuous-variable quantum-information.

The ten generators separate into two classes, corresponding to
noncompact and compact directions in ${\rm sp}(4,{\mathbb R})$.  The
noncompact (boost-like) generators are squeezing transformations, including
single-mode and two-mode squeezing.  The quantities \(s_1\) and \(s_2\)
measure the local widths, of \(x_1\) and \(x_2\),
while their conjugate variables \(p_{s_1}\) and \(p_{s_2}\) generate
single-mode squeezing. The parameter \(\beta\) acts as a hyperbolic angle,
mapped to the range \(0< \beta < \pi\), and quantifies symmetric correlations
between the modes. Its conjugate momentum
\(p_\beta\) generates the corresponding two-mode squeezing dynamics.

The compact (rotation-like) generators describe phase and mode-mixing transformations.
The parameter \(\alpha\) defines a relative phase in the momentum-space correlations, while its conjugate momentum \(p_\alpha\) generates the associated correlation dynamics.

\begin{figure}
 \begin{center}
   \includegraphics[width=12cm]{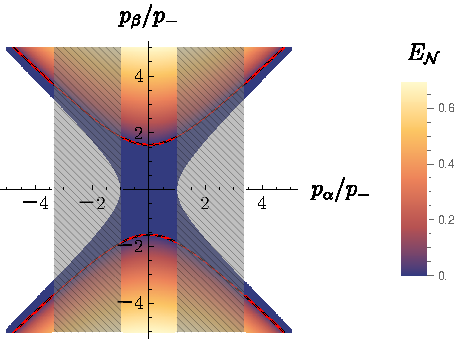}
  \caption{Logarithmic negativity ($\beta=\pi/2$, $C_1=2.7\hbar$, $C_2=2\hbar$)
is a function of the Minkowski distance of $(p_{\alpha},p_{\beta})$, as shown by
hyperbolic structures. Gray regions are excluded by positivity conditions on
(\ref{P}) and have boundaries represented by Riemannian geometry on extended
space-time, while it is Finslerian in the exteriors. The red dashed hyperbolae
where
$p_{\beta}^2-p_{\alpha}^2=\frac{1}{4}(\nu_-^2-\hbar^2/4)(\nu_+^2-\hbar^2/4)$
separate entangled states (top and bottom) from separable states
(interior). These hyperbolae turn into the light cone when $\nu_-$ takes its
minimum value, $\hbar/2$.  \label{fig:En}}
\end{center}
\end{figure}

The correlation dynamics generated jointly by \(p_\alpha\) and $p_{\beta}$ can be neatly summarized by expressing the
logarithmic negativity $E_{\cal N}$, derived from the covariance matrix as an
indicator of mode-entanglement as a function of the canonical variables. For $\beta=\pi/2$, the general expression takes the
manageable form
\begin{eqnarray} 2e^{-2E_{\cal
N}}|_{\beta=\pi/2}&=&C_1^2+4p_{\beta}^2-4p_{\alpha}^2\\ &&-
\sqrt{(C_1^2+4p_{\beta}^2-4p_{\alpha}^2)^2- C_1^4+C_2^4}\,. \nonumber
\end{eqnarray} This result, illustrated by the contour plot in
Fig.~\ref{fig:En}, shows that the Minkowski distance $p_{\beta}^2-p_{\alpha}^2$
is a dynamical measure of entanglement since the remaining contributions to
$E_{\cal N}$ are conserved by unitary evolution at the level of second-order
moments.

In the specific parameterization adopted here, the squeezing operations
generated by \(p_\beta\) and \(p_\alpha\) are not identical to those produced
by the standard two-mode squeezing operator \cite{GaussianStatesOperations},
$ \hat{S}_2(\xi) = \exp{(\xi^* \hat{a}_1 \hat{a}_2 - \xi \hat{a}_1^\dagger
  \hat{a}_2^\dagger)}$ with the complex parametrization
\(\xi = r \exp{i \theta}\) and mode ladder operators
\(\hat{a}_i, \hat{a}_i^\dagger\) for \(i = 1,2\).  The variables used here are
constructed by aligning the \(\beta\)-sector with the position block of the
covariance matrix, rather than with the ladder-operator basis. As shown in the Appendix, \(\beta = \pi/2\) corresponds to a
weakly squeezed configuration with vanishing correlation,
\(\Delta(x_1x_2) = 0\), whereas the limits \(\beta \rightarrow 0\) and
\(\beta \rightarrow \pi\) correspond to maximally correlated, Einstein-Podolsky-Rosen (EPR)-like
squeezed
states.

The remaining generators are the Lie-algebra Casimirs \(C_1\) and \(C_2\) and
represent conserved quantities of unitary two-mode dynamics. They are directly
related to the two independent symplectic eigenvalues \(\nu_\pm\) of the
two-mode covariance matrix by
\begin{equation} \label{CasimirsToEigenvalues}
C_1^2 = \nu_+^2 + \nu_-^2\quad,
\quad
C_2^2 = \nu_+^2 - \nu_-^2.
\end{equation}
The eigenvalues satisfy the uncertainty constraint $\nu_+\geq\nu_-\geq \hbar/2$ \cite{QuantumContinuous}, implying \(C_1 \ge \hbar /2\) and $C_2\geq 0$.
The limiting values, together with
$p_{\alpha}=0$, are achieved for a pure two-mode Gaussian state.
We therefore interpret \(p_\alpha\) as a pure-Gaussian witness: any nonzero value certifies that the state is not a pure Gaussian.
However, vanishing \(p_\alpha\) is not a sufficient condition for Gaussianity in the mixed-state sector.

In the language of quantum optics, the symplectic eigenvalues quantify the irreducible quantum noise associated with the two principal modes. Here, their significance extends beyond noise: they determine global information measures of the state, namely the entropy and purity.
The symplectic eigenvalues relate to the von Neumann entropy through
$S = s(2\nu_+/\hbar) + s(2\nu_-/\hbar)$ where
\begin{equation}
  s(\nu) = \frac{\nu+1}{2} \log_2\left( \frac{\nu + 1}{2}\right)
- \frac{\nu - 1 }{2} \log_2 \left( \frac{\nu - 1}{2}\right)\,,
\end{equation} and to the purity by
\begin{equation} \label{Purity} \mu =
\frac{\hbar^2/4}{\nu_+\nu_-}=\frac{\hbar^2/2}{\sqrt{C_1^4-C_2^4}}\,.
\end{equation}
Together with Eq.~(\ref{CasimirsToEigenvalues}), these relations establish a
direct correspondence between the geometric invariants \((C_1,C_2)\) and the
information-theoretic properties of the relativistic quantum system.

\section{Non-classical geometry}

Substituting the moment equations into the
time-dilation formula (\ref{ProperTime}) exposes a departure from classical
relativistic dynamics. In the classical case, the radicand is quadratic in the
momenta, reflecting the quadratic form of the Hamiltonian and the associated
Riemannian geometry.  By contrast, the quasiclassical Hamiltonian acquires a
non-polynomial dependence on the correlation momentum $p_{\alpha}$, which
enters the momentum moments through
\begin{equation} \label{P} \sqrt{P}=\sqrt{C_2^4-C_1^4+(C_1^2-4p_{\alpha}^2)^2}\,.
\end{equation}
The requirement that \(P\) remain real imposes a restriction
on the allowed values of \(p_\alpha\), excluding the range $p_-\leq
|p_{\alpha}|\leq p_+$ with $p_{\pm}=\frac{1}{2}(\nu_+\pm\nu_-)$. Outside of
this range, the effective metric is real, and quantum correlations
parameterized by \(p_\alpha\) obey  the uncertainty principle.

The expression (\ref{P}) is polynomial only for
special values of $C_1$ and $C_2$, such as when they are equal. However, this
case is forbidden by the fact that purity (\ref{Purity}) is bounded from above.
Therefore, the radicand of the time-dilation formula is non-quadratic in momenta
for non-pure non-Gaussian states ($p_{\alpha}\not=0$), and the geometry is non-Riemannian.
Since the radicand (\ref{P}) remains homogeneous of degree two in the momenta, the dynamics are governed by Finsler geometry \cite{FinslerDispersion,FinslerGeometryDispersion,FinslerGravity,FinslerGravityVariation}.

Our construction provides an explicit realization of the long-anticipated idea
that the interplay between gravity and quantum mechanics leads to new,
non-Riemannian geometric structures.  The construction here follows directly
from a canonical quantization of gravitationally coupled classical modes and
requires no additional hypotheses about the microscopic structure of quantum
space-time.  It relies only on the minimal assumption of geodesic motion
combined with the standard structure of quadrature moments familiar from
quantum optics.

For photons, entanglement
properties encoded in $p_{\alpha}$ imply features of non-linear optics even in
vacuum because the group velocity $\partial \langle\hat{E}\rangle/\partial
p_{\alpha}$ is a non-linear function of $p_{\alpha}$. Relevant experimental
constraints have been given in
\cite{DispersionConstraints}. Related observational effects include changes
to time dilation (\ref{ProperTime}), similar to
\cite{ProperTimeInterfer,ProperTimeInterfer2,ClockSystemInt,ClassQuantProperTime}
but here with explicit dependence on characteristic quantum-information
parameters and the underlying Finsler geometry of state space. As in these cases, the magnitude and detectability of these effects are governed by the momentum variances, and our results enable these contributions to be separated according to entropy and purity.

\section{Quantum contributions to time dilation}

An explicit formula that
determines implications of entropy and purity in a relativistic setting is given
by time dilation derived from our quantum Hamiltonian.
In flat space-time, the quantum-corrected Hamiltonian constraint entering Eq.~(\ref{ProperTime}) leads to the modified relation
\begin{equation}
  \label{eq:time-dilation-flat}
  \frac{{\rm d}t}{{\rm d}\tau}
  =
  \sqrt{
    1
  +
  \frac{
    g^{ij}(p_ip_j + \Delta(p_i p_j))
  }
  {m^2c^2}
  }\,,
\end{equation}
which shows that even for an object at rest, time dilation is implied by quantum fluctuations encoded in the momentum covariances 
\(\Delta(p_ip_j)\).

(Such an equation requires a Legendre transformation from the Hamiltonian to a
Lagrangian in order to bring in velocity terms such as ${\rm d}t/{\rm d}\tau$
replacing momenta. Given our non-polynomial Hamiltonian, a Legendre
transformation is non-trivial. It is possible to treat the contribution of the
square root (\ref{P}) as a small correction to a quadratic Hamiltonian and
then perform a perturbation expansion. However, the zeroth-order
transformation is then singular because $C_2$, which plays the role of a
conserved momentum, appears only in (\ref{P}) and not in the quadratic
contribution to the Hamiltonian.  These difficulties require a longer
derivation, for instance making use of the theory of constrained
systems. Here, we present the main result relevant for time dilation: Assuming
a non-moving particle in flat space-time, such that classical special and
general relativity would not imply time dilation, we obtain Even a stationary
clock not subject to a gravitational potential therefore experiences time
dilation. The expression given here in terms of symplectic eigenvalues shows
that quantum effects indeed imply time dilation, rather than a speed-up of
time.)

Equation~(\ref{eq:time-dilation-flat}) presents the contribution of momentum
fluctuations in an intuitive way. In flat space-time, only the variances
\(\Delta(p_i^2)\) contribute, but expressing $\dot{t}$ solely in terms of
these quantities hides much of the state's underlying geometric structure:
Within a two-mode Gaussian system, the covariance matrix satisfies positive
semidefiniteness and symplectic constraints, which couple all variances and
correlations.  Therefore, the momentum variances are not independent
coordinates of the state.  Furthermore, unlike in the classical limit where
the identification \(p/m \leftrightarrow \dot{x}\) follows directly from the
Lagrangian, the quantum momenta acquire geometric meaning only after a
canonical mapping that embeds the statistical moments into an effective
phase-space.  (The canonical transformation provides coordinates on the
symplectic manifold of allowed covariances which factorize the uncertainty
constraints. Each coordinate can vary freely within simple domains
(\(\mathbb{R}\) or bounded intervals) while still preserving the symplectic
condition.  By contrast, working directly with \(\Delta(p_1^2)\) and
\(\Delta(p_2^2)\) requires keeping track of nonlinear algebraic constraints
that keep the whole covariance matrix physically valid.)

We substitute the canonical map for the momentum variances, obtaining a
function $\dot{t}$ that is convex in the quantum momenta \(p_{s_1}, p_{s_2},
p_\alpha, p_\beta\). To isolate the role of correlations and fundamental noise
($\nu_{\pm}$), consider an instant where the canonical momenta vanish,
\begin{equation}
  p_{s_1} = p_{s_2} = p_\alpha = p_\beta = 0
\end{equation}
corresponding to the minimum of the convex function
(\ref{eq:time-dilation-flat}) on the quantum phase space.  The remaining
dependence enters only through the symplectic eigenvalues \(\nu_\pm\), which
obey \(\nu_\pm \ge \hbar/2\) and are fixed by the initial state in Hamiltonian
evolution. (They change only under non-Hamiltonian processes such as loss,
dephasing, or active cooling/amplification.)

In this regime the two-mode time-dilation factor depends on the residual noise carried by the principal modes, weighted by the correlation geometry
  \begin{equation} \left(\frac{{\rm d}t}{{\rm d}\tau}\right)^2=
      1
  +
  \frac{
    A_+(s_1, s_2, \alpha, \beta) \, \nu_+^2
    +
    A_-(s_1, s_2, \alpha, \beta) \, \nu_-^2
  }{2m^2c^2s_1^2s_2^2\sin^2\beta}
\end{equation}
with
\begin{equation}
  \label{eq:19}
  A_{\pm} =
          s_1^2 \left(1 \mp \sin(\alpha - \beta) \right)
          +
          s_2^2 \left(1 \mp \sin(\alpha + \beta) \right)\,.
\end{equation}
The overall scale of the effect is set by state entropy and purity through the symplectic eigenvalues
\(\nu_\pm\). Greater mixedness (larger \(\nu_\pm\)) amplifies the time-dilation penalty, while purer states (smaller 
\(\nu_\pm\), bounded below by \(\hbar/2\)) suppress it. The correlation angles \((\alpha,\beta)\) and the lever arms 
\((s_1,s_2)\) determine how this intrinsic quantum noise is partitioned into the observable dilation. 
The weights also depend on the gravitational potential if
the metric components from (\ref{eq:time-dilation-flat}) are included.

As an application to precision time-keeping, note that steering the
correlation parameters \(\alpha\) and \(\beta\) allows one to redistribute
noise between the two symplectic channels \((\nu_+, \nu_-)\).  For
approximately symmetric widths \(s_1 \approx s_2 = s\), the time-dilation
factor simplifies to
\begin{equation} \label{symmetric-dtdtau}
\frac{{\rm d}t}{{\rm d}\tau}
\approx
  1
  +
  \frac{
    (1 - \sin\alpha\cos\beta) \, \nu_+^2
    +
    (1 + \sin\alpha\cos\beta) \, \nu_-^2
  }{2m^2c^2s^2\sin^2\beta}\,.
\end{equation}
Selecting \(\alpha\) such that the larger symplectic eigenvalue multiplies the smaller coefficient minimizes the effective dilation.
In the ideal limit \(\vert \sin \alpha \cos\beta \vert = 1\), the contribution from the noisier channel cancels entirely, leaving the time-dilation rate determined only by the purer mode. This extreme suppression is achievable only near maximal inter-mode coupling, \(\beta \approx 0\) or \( \beta \approx \pi\), indicating that entanglement becomes a practical resource for high-precision quantum timekeeping.
In this limit the denominator of Eq.~(\ref{symmetric-dtdtau}) tends to zero, reflecting the divergence of momentum fluctuations required by symplectic volume conservation. The overall effect is an increase in the effective time-dilation rate as the state approaches the EPR limit, illustrating that optimal clock performance requires balancing correlations by jointly tuning \(\alpha\) and \(\beta\) for given symplectic spectra \(\nu_\pm\).

Correlations steer the phase-space orientation of noise; entanglement enables
selective noise suppression in the proper-time rate.  As an estimate of the
order of magnitude, we can set $\beta=\pi/2$ and assume $s:=s_1\approx s_2$
and that the larger symplectic eigenvalue $\nu_+$ dominates. Both $\alpha$ and
$\beta$ then disappear from the equation, and we obtain
 \begin{equation} \frac{{\rm d}\tau}{{\rm d}t} \approx 1-\frac{\nu_+^2}{2s^2
 m^2c^2}=1-\frac{\lambda^2}{8\pi^2s^2} \frac{\nu_+^2}{\hbar^2}\,.
\end{equation} In the final step, we used the Compton wave length
$\lambda=h/(mc)$ of the clock and the dimensionless ratio $\nu_+/\hbar$.

\section{The weight of purity}

Additional implications can be seen if we
evaluate quantum geodesic motion on curved space-time, in the presence of a
gravitational force. The quasiclassical Hamiltonian then implies gravitational
effects of fluctuation terms $g^{ab}\Delta(p_ap_b)$ that can be interpreted as
a contribution to the effective mass
\begin{equation} \label{meff}
  m_{\rm
    eff}=\sqrt{m^2+g^{ab}\Delta(p_ap_b)/c^2}\,.
\end{equation}
Even a classical massless object (light or a gravitational wave)
therefore acquires non-zero effective mass from non-vanishing momentum
fluctuations.  Through the square root $P$, entropy and purity contribute to the
effective mass. In particular, more purity (\ref{Purity}) implies a smaller
value of $C_1^4-C_2^4$, which is subtracted in the radicand of $P$. Thus, $P$
grows for larger purity. Combining all $P$-terms in the moments, this expression
appears with a coefficient whose sign is determined by
\begin{equation}
  -g^{11}\frac{\sin(\alpha+\beta)}{s_1^2}+g^{12}\frac{\sin\alpha}{s_1s_2}
  -g^{22}\frac{\sin(\alpha-\beta)}{s_2^2}
\end{equation}
and may be positive or negative. Upon using trigonometric identities, this
expression can be split in the isotropic contribution
$-(g^{11}/s_1^{2}+g^{22}/s_2^2)\sin\alpha\cos\beta$ proportional to quantum
correlations, and an anisotropic term that vanishes for isotropic configurations
as defined by $g^{12}=0$ and $g^{11}=g^{22}$ for the metric, and $s_1=s_2$ for
the state.

Depending on the two correlation parameters $\alpha$ and $\beta$, the weight
of a quantum object may be increased or decreased by its purity, compared with
a pure Gaussian state for which $P=0$. Using our new methods, it is possible
to evaluate such implications for quantum states in free fall or in an
electromagnetic trap. The same methods can also be applied to astrophysical
situations, such as photons deflected or emitted around heavy masses, or to
quantum gravitational waves.

An anisotropic example is given by equatorial motion, such that
$g^{11}=1-2GM/(c^2r)$ in the radial direction of Schwarzschild space-time
surrounding a central mass $M$, while $g^{22}=r^{-2}$ does not depend on
$M$. Through $P$, purity then modifies Newton's constant via the effective
mass. Compared with effective field theory
\cite{EffectiveNewton,EffectiveGR,BurgessLivRev,LongRangeQG,QGNewton,NewtonOneLoop,TwoMasses,NewtonKerr,EffNewtonCoulomb},
the coefficients have a specific quantum-information content and allow more
sensitive tests.

Another implication is that quantum geodesic motion can be used to test
properties of Hawking radiation.  The derivation given here can be seen as a
geometric-optics approximation to quantum-field theory on a curved
background. Our methods make it easier to include direct effects from
quantum-information properties of the Hawking state, including radiation from
rotating black holes through quantum geodesic motion on a Kerr space-time.

The effective mass (\ref{meff}) is a useful tool to discuss qualitative
implications. A similar expression applies generically to classically massless
objects, which acquire an effective mass through momentum fluctuations. One
might worry that the effective mass would lead to recapture of Hawking photons
by the black hole and strongly modify the traditional spectrum as well as
evaporation rate. However, the quasiclassical energy simplifies for radial
motion of Hawking photons escaping from a non-rotating black hole: In this
case, the relevant classical energy expression $E$ in (\ref{Hexp}) is linear
in the only non-zero momentum, $p_r$, such that the quasiclassical energy is
not amended by momentum fluctuations. There may be position fluctuations as
well as position-momentum covariances since the metric component $g^{rr}$ is
position-dependent. But the former would only modify the coefficient of $p_r$
in the energy, while the latter can be assumed small or strictly
vanishing. The radial case of massless objects therefore does not imply an
effective mass, and therefore respects standard properties of Hawking
radiation.  Non-trivial effects are expected for non-radial motion, as in
Hawking radiation around a rotating black hole or in lightlike geodesics that
form the photon ring around a black hole.

\section{Summary and outlook}

A canonical parametrization of second-order
moments provides a geometric representation of quantum fluctuations and
reveals their direct quantum-informational meaning --- encoding entropy, purity
and correlations through symplectic invariants that appear naturally in the
formalism. Detailed constructions allowed
us to suggest control mechanisms for time dilation, and they can be applied to
other relativistic effects such as tidal forces, and geodesic deflection. The
canonical embedding of the covariance matrix into an effective phase space
makes transparent which squeezing transformations are genuinely distinct
(non-passive), and which are related by local symplectic rotations. The same
invariants that define this symplectic embedding determine the state's entropy
and purity.  This duality between geometry and information represents a key
conceptual result: it provides a setting in which quantum information and
relativistic geometry coexist within a common canonical formalism.

As a new feature of quantum space-time, we have shown from first principles that a subtle quantum parameter representing a new type of inter-mode correlation is naturally associated with a non-Riemannian (Finsler-like) geometry. This result suggests that even in flat space-time, quantum correlations can induce geometric structures beyond the classical metric description.

Experimentally, such correlations may be accessible in spin-squeezed states of
trapped ions or cold atomic ensembles, which already demonstrate reduced
quantum-projection noise in phase measurements
\cite{schulte_prospects_2020,pezze_quantum_2018,
  hosten_measurement_2016,leroux_orientation-dependent_2010}. Existing
studies, however, have not yet recognized the phase-space geometry
underlying these effects. Recognizing that structure could open new avenues
for controlling geometric and informational aspects of quantum states,
potentially leading to improved entanglement-assisted timekeeping, refined
models of quantum-gravitational redshift, or even tests of non-Riemannian
signatures in engineered quantum systems.

\section*{Acknowledgements}

\noindent This work was supported in part by NSF Grant No. PHY-2206591.


\begin{thebibliography}{10}

  \bibitem{QuantumInfoRel}
A. Peres and D. Terno, Rev.\ Mod.\ Phys. {\bf 76},  93  (2004),
  quant-ph/0212023.

\bibitem{QuantumFinsler}
J. Balsells and M. Bojowald, Phys.\ Rev.\ D {\bf 113}, 124060 (2026),  arXiv:2503.06667.

\bibitem{Bosonize}
B. Bayta\c{s}, M. Bojowald, and S. Crowe, Ann.\ Phys. {\bf 420},  168247
  (2020), arXiv:1810.12127.

\bibitem{EffPotRealize}
B. Bayta\c{s}, M. Bojowald, and S. Crowe, Phys.\ Rev.\ A {\bf 99},  042114
  (2019), arXiv:1811.00505.

\bibitem{GaussianStatesOperations}
J.~B. Brask, Gaussian states and operations -- a quick reference,
  arXiv:2102.05748.

\bibitem{QuantumContinuous}
A. Serafini, {\em Quantum continuous variables} (CRC Press, London, England,
  2021).

\bibitem{FinslerDispersion}
F. Girelli, S. Liberati, and L. Sindoni, Phys.\ Rev.\ D {\bf 75},  064015
  (2007), gr-qc/0611024.

\bibitem{FinslerGeometryDispersion}
D. Raetzel, S. Rivera, and F.~P. Schuller, Phys.\ Rev.\ D {\bf 83},  044047
  (2011), arXiv:1010.1369.

\bibitem{FinslerGravity}
C. Pfeifer and M.~N.~R. Wohlfarth, Phys.\ Rev.\ D {\bf 85},  064009  (2012),
  arXiv:1112.5641.

\bibitem{FinslerGravityVariation}
M. Hohmann, C. Pfeifer, and N. Voicu, Phys.\ Rev.\ D {\bf 100},  064035
  (2019), arXiv:1812.11161.

\bibitem{DispersionConstraints}
G. Amelino-Camelia, C. L\"ammerzahl, F. Mercati, and G.~M. Tino, Phys.\ Rev.\
  Lett. {\bf 103},  171302  (2009), arXiv:0911.1020.

\bibitem{ProperTimeInterfer}
M. Zych, F. Costa, I. Pikovski, and C. Brukner, Nat.\ Commun. {\bf 2},  505
  (2011), arXiv:1105.4531.

\bibitem{ProperTimeInterfer2}
M. Zych {\it et~al.}, Class.\ Quantum Grav. {\bf 29},  224010  (2012),
  arXiv:1206.0965.

\bibitem{ClockSystemInt}
A.~R.~H. Smith and M. Ahmadi, Quantum {\bf 3},  160  (2019), arXiv:1712.00081.

\bibitem{ClassQuantProperTime}
A.~R.~H. Smith and M. Ahmadi, Nat.\ Commun. {\bf 11},  5360  (2020),
  arXiv:1904.12390.

\bibitem{EffectiveNewton}
J.~F. Donoghue, Phys.\ Rev.\ Lett. {\bf 72},  2996  (1994), gr-qc/9310024.

\bibitem{EffectiveGR}
J.~F. Donoghue, Phys.\ Rev.\ D {\bf 50},  3874  (1994), gr-qc/9405057.

\bibitem{BurgessLivRev}
C.~P. Burgess, Living Rev.\ Relativity {\bf 7},  5  (2004),
  gr-qc/0311082.http://www.livingreviews.org/lrr-2004-5.

\bibitem{LongRangeQG}
I.~J. Muzinich and S. Vokos, Phys.\ Rev.\ D {\bf 52},  3472  (1995),
  hep-th/9501083.

\bibitem{QGNewton}
H.~W. Hamber and S. Liu, Phys.\ Lett.\ B {\bf 357},  51  (1995),
  hep-th/9505182.

\bibitem{NewtonOneLoop}
A. Akhundov, S. Bellucci, and A. Shiekh, Phys.\ Lett.\ B {\bf 395},  16
  (1997), gr-qc/9611018.

\bibitem{TwoMasses}
N.~E.~J. Bjerrum-Bohr, J.~F. Donoghue, and B.~R. Holstein, Phys.\ Rev.\ D {\bf
  67},  084033  (2003), hep-th/0211072.

\bibitem{NewtonKerr}
N.~E.~J. Bjerrum-Bohr, J.~F. Donoghue, and B.~R. Holstein, Phys.\ Rev.\ D {\bf
  68},  084005  (2003), hep-th/0211071.

\bibitem{EffNewtonCoulomb}
S. Faller, Phys.\ Rev.\ D {\bf 77},  124039  (2008), arXiv:0708.1701.

\bibitem{schulte_prospects_2020}
M. Schulte {\it et~al.}, Nature Communications {\bf 11},  5995  (2020).

\bibitem{pezze_quantum_2018}
L. Pezz\`e {\it et~al.}, Rev.~Mod.~Phys. {\bf 90},  035005  (2018).

\bibitem{hosten_measurement_2016}
O. Hosten, N.~J. Engelsen, R. Krishnakumar, and M.~A. Kasevich, Nature {\bf
  529},  505  (2016).

\bibitem{leroux_orientation-dependent_2010}
I.~D. Leroux, M.~H. Schleier-Smith, and V. Vuleti\'c, Phys.~Rev.~Lett. {\bf
  104},  250801  (2010).

\end{thebibliography}

\begin{appendix}
\section{Canonical moments}

The complete set of second-order moments in terms of
canonical variables:
  \begin{eqnarray} \Delta(x_1^2)&=&s_1^2 \quad,\quad \Delta (x_2^2)=s_2^2 \\
\Delta(x_1p_1)&=&s_1p_{s_1}\quad,\quad \Delta(x_2p_2)=s_2p_{s_2}\\
\Delta(x_1x_2)&=& s_1s_2\cos\beta\\ \Delta(x_1p_2) &=&s_1 p_{s_2} \cos(\beta) -
\sin(\beta)\frac{s_1}{s_2}\left(p_{\alpha}+p_{\beta}\right)\\ \Delta(x_2p_1)
&=&s_2 p_{s_1}
\cos(\beta)+\sin(\beta)\frac{s_2}{s_1}\left(p_{\alpha}-p_{\beta}\right)
\end{eqnarray} 
  \begin{equation} \label{p1app}
    \Delta(p_1^2)
    = p_{s_1}^2
    +
    \frac{(p_{\alpha}-p_{\beta})^2}{s_1^2}
    + \frac{1}{2
      s_1^2\sin^2{(\beta)}}
    \left(
      C_1^2-4 p_{\alpha}^2
      - \sqrt{C_2^4-C_1^4+(C_1^2-4
        p_{\alpha}^2)^2}\sin{(\alpha+\beta)}
    \right)
\end{equation}
\begin{equation} \label{p2}
  \Delta(p_2^2)
  =
  p_{s_2}^2
  +
  \frac{(p_{\alpha}+p_{\beta})^2}{s_2^2}
  + \frac{1}{2
s_2^2\sin^2(\beta)}\left(C_1^2-4 p_{\alpha}^2-\sqrt{C_2^4-C_1^4+(C_1^2-4
p_{\alpha}^2)^2}\sin(\alpha-\beta)\right)
\end{equation} and
\begin{eqnarray}
  \label{eq:Deltap1p2} \Delta(p_1 p_2)
  &=&
      \left(
      p_{s_1}p_{s_2}
      +
      \frac{p_{\alpha}^2- p_{\beta}^2}{s_1s_2}
      \right) \cos(\beta)
     +
     \left[
     p_{\alpha}
     \left(\frac{p_{s_2}}{s_1}-\frac{p_{s_1}}{s_2}\right)
     -
     p_{\beta}
     \left(\frac{p_{s_2}}{s_1}
     +
     \frac{p_{s_1}}{s_2}\right)
     \right] \sin(\beta)
\nonumber     \\
  &&
     -
     \frac{1}{2s_1s_2\sin^2(\beta)}
     \left(
     (C_1^2 - 4p_{\alpha}^2) \cos(\beta)
     - \sqrt{C_2^4 - C_1^4 +
     (C_1^2 - 4p_{\alpha}^2)^2} \,\sin(\alpha)
     \right)
     \,. \nonumber
\end{eqnarray}
The paper \cite{QuantumFinsler} provides the inverse map, expressing all
canonical variables directly in terms of second central moments.

\end{appendix}

\end{document}